\pdfoutput=1
\documentclass{article}
\usepackage{ijcai19}
\usepackage{booktabs} 
\usepackage{amsmath}
\usepackage{amssymb}
\usepackage[ruled, linesnumbered]{algorithm2e}
\usepackage{amsmath}
\usepackage{xcolor}
\usepackage{multirow}
\usepackage{graphicx}
\usepackage{subfigure}
\usepackage{algorithmic}
\usepackage{amsfonts}
\usepackage{titlesec}

\newtheorem{definition}{Definition}

\vspace{-0.3cm}

\title{Personalized Multimedia Item and Key Frame Recommendation}

\vspace{-0.3cm}
\author{
Le Wu$^1$\and
Lei Chen$^1$\and
Yonghui Yang$^1$\and
Richang Hong$^1$\and
\\
Yong Ge$^2$\and
Xing Xie$^3$\and 
Meng Wang$^1$ 
\affiliations
$^1$Hefei University of Technology\\
$^2$The University of Arizona\\
$^3$Microsoft Research 
\emails
\{lewu.ustc,chenlei.hfut, yyh.hfut,hongrc.hfut\}@gmail.com,  
yongge@email.arizona.edu,
xingx@microsoft.com,      
eric.mengwang@gmail.com 
}

\begin{document}

\vspace{-0.3cm}
\maketitle

\vspace{-1.0cm}
\begin{abstract}
When recommending or advertising items to users, an emerging trend is to present each multimedia item with  a \emph{key frame} image~(e.g., the poster of a movie). As each multimedia item can be represented as  multiple fine-grained  visual images~(e.g., related images of the movie), personalized key frame recommendation is necessary in these applications to attract users' unique visual preferences. However, previous personalized key frame recommendation models relied on users' fine-grained image  behavior of  multimedia items~(e.g., user-image interaction behavior), which is often not available in real scenarios.  In this paper, we study the general problem of joint multimedia item and key frame recommendation in the absence of the fine-grained user-image behavior. We argue that the key challenge of this problem lies in discovering users' visual profiles for key frame recommendation, as most recommendation models  would fail without any users' fine-grained image behavior. To tackle this challenge, we leverage users' item behavior by projecting users~(items) in two latent spaces: a collaborative latent space and a visual latent space. We further design a model to discern both the collaborative and  visual dimensions of users, and model how users make decisive item preferences from these two spaces. As a result, the learned user visual profiles could be directly applied for key frame recommendation. Finally, experimental results on a real-world dataset clearly show the effectiveness of our proposed model on the two recommendation tasks.
\end{abstract}

\vspace{-0.1cm}
\section{Introduction}

Living in a digital world with overwhelming information, the visual based content, e.g., pictures, and images, is usually the most eye-catching for users and convey specific views to users~\cite{zhang2018image,yin2018transcribing,chen2019quality}.  Therefore, when recommending or advertising multimedia items to users, an emerging trend is to present each multimedia item with a display image, which we call the \emph{key frame} in this paper. E.g., as shown in Fig.~\ref{fig:example}, the movie recommendation page usually displays  each movie with a poster to attract users' attention. Similarly, for (short) video recommendation, the title cover is directly presented to users to quickly spot the visual content. Besides, image-based advertising promotes the advertising item with an attractive image, with 80\% of marketers use visual assets in the social media marketing~\cite{stelzner2018}.

\begin{small}
\begin{figure*} [htb]
 \begin{center}
\includegraphics[width=150mm]{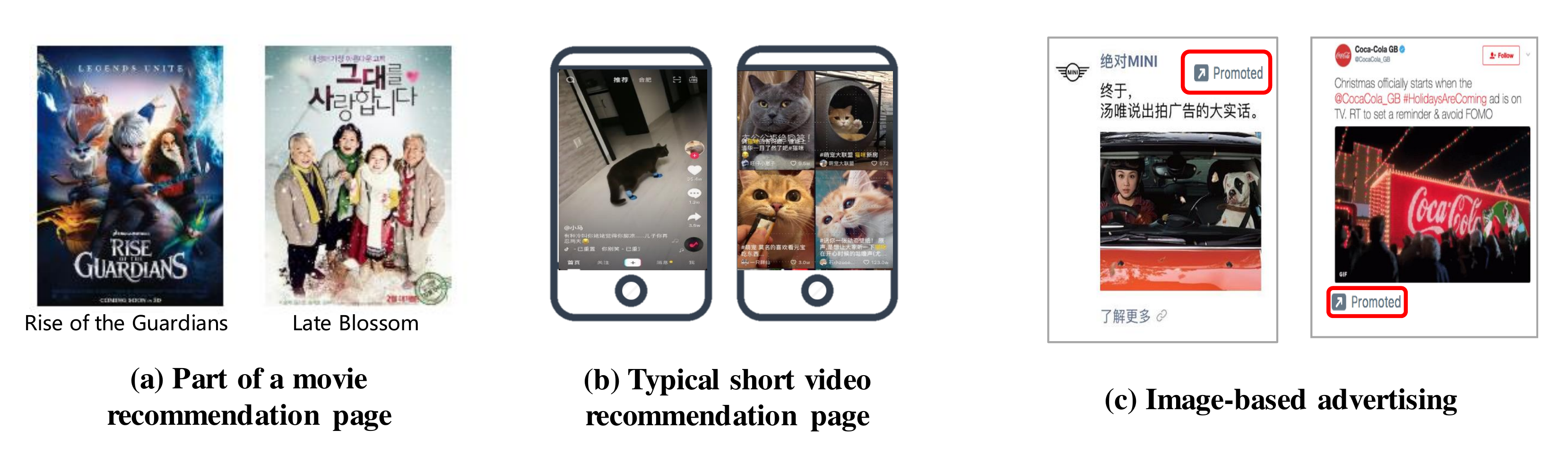}
  \end{center}
  \vspace{-0.7cm}
  \caption{\small{Three application scenarios with the key frame presentation of a multimedia item.} } \label{fig:example}
  \vspace{-0.6cm}
\end{figure*}
\end{small}

As a key frame distills the visual essence of a multimedia item, key frame extraction and recommendation for multimedia items  is widely investigated in the past ~\cite{IJDI2006keyframe,CVPR2013video,ICCV2011key}. Some researchers focused on how to summarize representative video content from videos, or apply image retrieval models from text descriptions for universal key frame extraction~\cite{IJDI2006keyframe,MM2014deep}. In the real-world, users' visual preferences are not the same but vary from person to person~\cite{pnas2017psychological,wu2019hierarchical}. By collecting users' behaviors to the images of the multimedia items, many recommendation models could be applied for personalized key frame recommendation~\cite{TKDE2005toward,UAI2009bpr,youtube2016deep}. Recently, researchers have made one of the first few attempts to design a computational KFR model that is tailored for personalized key frame recommendation in videos~\cite{sigir2017key}. By exploring the time-synchronized comment behavior from video sharing platforms, the authors proposed to encode both  users' interests and the visual and textual features of the frame in a unified multi-modal space to enhance key frame recommendation performance.

Despite the preliminary attempts for personalized key frame recommendation, we argue that these models not well designed due to the following two reasons. First, as the key frame is a visual essence of a multimedia item, recommending a key frame is always associated with the corresponding recommended multimedia item.  Therefore, is it possible to design a model that simultaneously recommends both items and  personalized key frames? Second,  most of the current key frame recommendation models relied on the fine-grained frame-level user preference in the modeling process, i.e., users' behaviors to the frames of the multimedia item, which is not available for most platforms.  Specifically, without any users' frame behavior, the classical collaborative filtering based models fail as these models rely on the user-frame behavior for recommendation~\cite{UAI2009bpr,koren2009matrix}. The content based models also could not work as these models need users' actions to learn the visual dimensions of users~\cite{IJCAI2018interpretable,CVPR2016comparative}. Therefore, the problem of how to build the user visual profile for key frame recommendation when the fine-grained user-image behavior data is not available remains challenging.

In this paper, we study the general problem of personalized multimedia item and key frame recommendation without fine-grained user-image interaction data. The key idea of our proposed model is to distinguish each user's latent collaborative preference and the visual preference from her multimedia item interaction history, such that each user's visual dimensions could be transferred for visual based key frame recommendation.  We design a \emph{J}oint \emph{I}tem and key \emph{F}rame \emph{R}ecommendation~(JIFR) model to  discern both the collaborative and  visual dimensions of users, and model how users make item preference decisions from these two spaces. Finally, extensive experimental results on a dataset clearly show the effectiveness of our proposed model for the two personalized recommendation tasks.

\vspace{-0.1cm}
\section{Problem Definition}

In a multimedia item recommender system, there are a userset  {\small$\mathcal{U}$~($|\mathcal{U}|\!=\!M$)}, and an itemset {\small$\mathcal{V}$~($|\mathcal{V}|\!=\!N$)}. Each multimedia item $v$ is composed of many \emph{frames}, where each frame is an image that shows a part of the visual content of this item.  Without confusion, we would use the term of frame and image interchangeably in this paper. E.g., in a movie recommender system, each movie is a multimedia item.  There are many images that could describe the visual content of this movie, e.g., the official posters in different countries, the trailer posters, and the frames contained in this video. For video recommendation,  it is hard to directly analyze the video content from frame to frame, a natural practice is to extract several typical frames that summarize the content of this video~\cite{IJDI2006keyframe}. Besides, for image-based advertising, for each advertising content with text descriptions, many related advertising images~(frames) could be retrieved with text based image retrieval techniques~\cite{MM2014deep}.  

Therefore, besides users and items, all the frames of the multimedia items in the itemset compose a frameset {\small$\mathcal{C}$~($|\mathcal{C}|\!=\!L$)}.  The relationships between items and frames  are represented as a \emph{frame-item correlation matrix} {\small$\mathbf{S}\in \mathbb{R}^{L\times N}$}.
We use {\small$S_v=[j| s_{jv}\!=\!1]$} to denote the frames of a multimedia item $v$. For each item $v$, the \emph{key frame} is the display image of this multimedia item when it is presented or recommended to users. Therefore, the key frame belongs to $v$'s frameset $S_v$. Besides, in the multimedia system, users usually show implicit feedbacks~(e.g., watching a movie, clicking an advertisement) to items, which could be represented as a \emph{user-item rating matrix} {\small$\mathbf{R}\in \mathbb{R}^{M\times N}$}. In this matrix, $r_{ai}$ equals 1 if $a$ shows preferences for item $i$, otherwise it equals 0.

\begin{definition}  \label{def:prob_def}[\textbf{Multimedia Item and Key Frame Recommendation}] In a multimedia recommender system, there are three kinds of entities: a userset {\small$\mathcal{U}$}, an itemset {\small$\mathcal{V}$}, and a frameset {\small$\mathcal{C}$}.  The item multimedia content is represented as a frame-item correlation matrix {\small$\mathbf{S}$}.  Users show item preference with user-item implicit rating matrix {\small$\mathbf{R}$}. For visual based multimedia recommendation, our goal is two-fold: 1) \emph{Multimedia Item Recommendation}: Predict each user $a$'s unknown preference $\hat{r}_{ai}$ to multimedia item $i$; 2) \emph{Key Frame Recommendation}: For user $a$ and the recommended multimedia item $i$, predict her unknown find-grained preference to multimedia content $\hat{l}_{ak}$ , where $k$ is a multimedia image content of $i$~($s_{ki}=1$).
\end{definition}

\section{The Proposed Model}

In this section, we introduce our proposed  JIFR model for multimedia item and key frame recommendation. We start with the overall intuition, followed by the detailed model architecture and the model training process.
At the end of this section, we analyze the proposed model.

With the absence of the fine-grained user behavior data, it is natural to leverage the user-item rating matrix {\small $\mathbf{R}$} to learn each user's profile for key frame recommendation. Therefore, in each user's decisive process for multimedia items, we adopt a hybrid recommendation model that projects users and multimedia items into two spaces: a latent space to characterize the collaborative behavior, and a visual space  that shows the visual dimensions that influences users' decisions. Let  {\small$\mathbf{U}\in \mathbb{R}^{d1\times M}$} and {\small$\mathbf{V}\in \mathbb{R}^{d1\times N}$} denote the free user and item latent vectors in the collaborative space, and {\small$\mathbf{W}\in \mathbb{R}^{d2\times M}$} and {\small$\mathbf{X}\in \mathbb{R}^{d2\times N}$} are the visual representations of users and items in the visual dimensions. For the visual dimension construction, as each multimedia item $i$ is composed of multiple images, its visual representation $\mathbf{x}_i$ is summarized from the related visual embeddings of the corresponding frame set $S_i$ as:

\begin{small}
\vspace{-0.4cm}
\begin{equation} \label{eq:vis_avg_i}
\mathbf{x}_i=\sum_{k\in S_i} \frac{\mathbf{P}\mathbf{c}_k}{|S_i|},
\end{equation}
\vspace{-0.2cm}
\end{small}

\noindent where $\mathbf{c}_k$ is the visual representation of image $k$. Due to the success of convolutional neural networks, similar as many visual modeling approaches, we use the last fully connected layer in VGG-19 to represent the visual content of each image $k$ as $\mathbf{c}_k\in\mathbb{R}^{4096}$~\cite{vgg2014very,aaai2016vbpr}. {\small $\mathbf{P}$}$\mathbf{c}_k$ transforms the original item visual content representation from 4096 dimensions into a low latent visual space, which is usually less than 100 dimensions.

Given the multimedia item representation, each user's predicted preference could be seen as a hybrid preference decision process that combines the collaborative filtering preference and the visual content preference as:

\begin{small}
\vspace{-0.2cm}
\begin{equation} \label{eq:pred_vbpr}
\hat{r}_{ai}=\mathbf{u}^T_a\mathbf{v}_i+\mathbf{w}_a^T\mathbf{x}_i,
\end{equation}
\vspace{-0.3cm}
\end{small}

\noindent  where $\mathbf{w}_a$ is the visual embedding of user $a$ from the user visual matrix {\small$\mathbf{W}$}.  In fact, by summarizing the item visual content from its related frames, the above equation is similar to the VBPR model that uncovers the visual and latent dimensions of users~\cite{aaai2016vbpr}. 

With the implicit feedbacks of the rating matrix {\small$\mathbf{R}$}, Bayesian Personalized Ranking~(BPR) is widely used for modeling the pair-wise based optimization function~\cite{UAI2009bpr}:

\begin{small}
 \vspace{-0.3cm}
\begin{equation}\label{eq:loss_vbpr}
\min\mathcal{L}_{R}=\sum_{a=1}^M\sum\limits_{(i,j)\in D^R_a }\sigma(\hat{r}_{ai}-\hat{r}_{aj}) +\lambda_1||\Theta_1||_F^2,
\end{equation}
 \vspace{-0.4cm}
\end{small}

\noindent where {\small$\Theta_1=[\mathbf{U},\mathbf{V},\mathbf{W}]$} is the parameter set, and $\sigma(x)$ is a sigmoid function that transforms the output into range $(0,1)$. The training data for user $a$ is {\small$D^R_a=\{(i,j)|i\in R_a\!\wedge\!j\in V-R_a\}$}, where {\small$R_a$} denotes the set of implicit positive feedbacks of $a$~(i.e., {\small$r_{ai}\!=\!1$}), and {\small$j\!\in\!V-R_a$} is an unobserved feedback.

In key frame decision process, each key frame image presentation of the current item  is the foremost visual impression to influence and persuade users. By borrowing the learned user visual representation matrix {\small$\mathbf{W}$} from the user-item interaction behavior~(Eq.\eqref{eq:loss_vbpr}), each user's visual preference for image $k$ is predicted as:

\begin{small}
\vspace{-0.2cm}
\begin{equation} \label{eq:pred_img}
\hat{l}_{ak}=\mathbf{w}^T_a(\mathbf{P}\mathbf{c}_k),
\end{equation}
\vspace{-0.3cm}
\end{small}

\noindent where $\mathbf{w}_a$ is the visual latent embedding of user $a$ learned from user-item interaction behavior. Please note that, the predicted $\hat{l}_{ak}$ is only used in the test data for the key frame recommendation without any training process.

Under the above approach, for each user, by optimizing the user-item based loss function~(Eq.\eqref{eq:loss_vbpr}), we could align users and images in the visual space without any user-image interaction data for multimedia item and key frame recommendation.  However, we argue that the above approach is not well designed for the proposed problem due to the overlook of the
\emph{content indecisiveness} and \emph{rating indecisiveness} in the modeling process. The content indecisiveness is  correlated to the visual representation of each item~(Eq.\eqref{eq:vis_avg_i}), which refers to which images are more important to reflect the visual content of the multimedia item are unknown. E.g., some frames in the movie convey more visual semantics than other frames that are not informative. Simply using an average operation that summarizes the item visual representation~(Eq.\eqref{eq:vis_avg_i}) would neglects the visual highlights of the item semantics. Besides, the rating indecisiveness appears in each user-item preference decision process as shown in Eq.\eqref{eq:pred_vbpr}, which refers to the implicitness of whether to concentrate more on the collaborative part or the visual item part for the preference decision process. For example, sometimes a user chooses a movie since this movie is  visually stunning, even though this movie does not follow her historical watching histories. Therefore, the recommendation performance is limited by the assumption that the hybrid preference is equally contributed by the collaborative and visual content based models as shown in Eq.\eqref{eq:pred_vbpr}.

\begin{small}
\begin{figure} [htb]
 \begin{center}
 \vspace{-0.2cm}
\includegraphics[width=88mm]{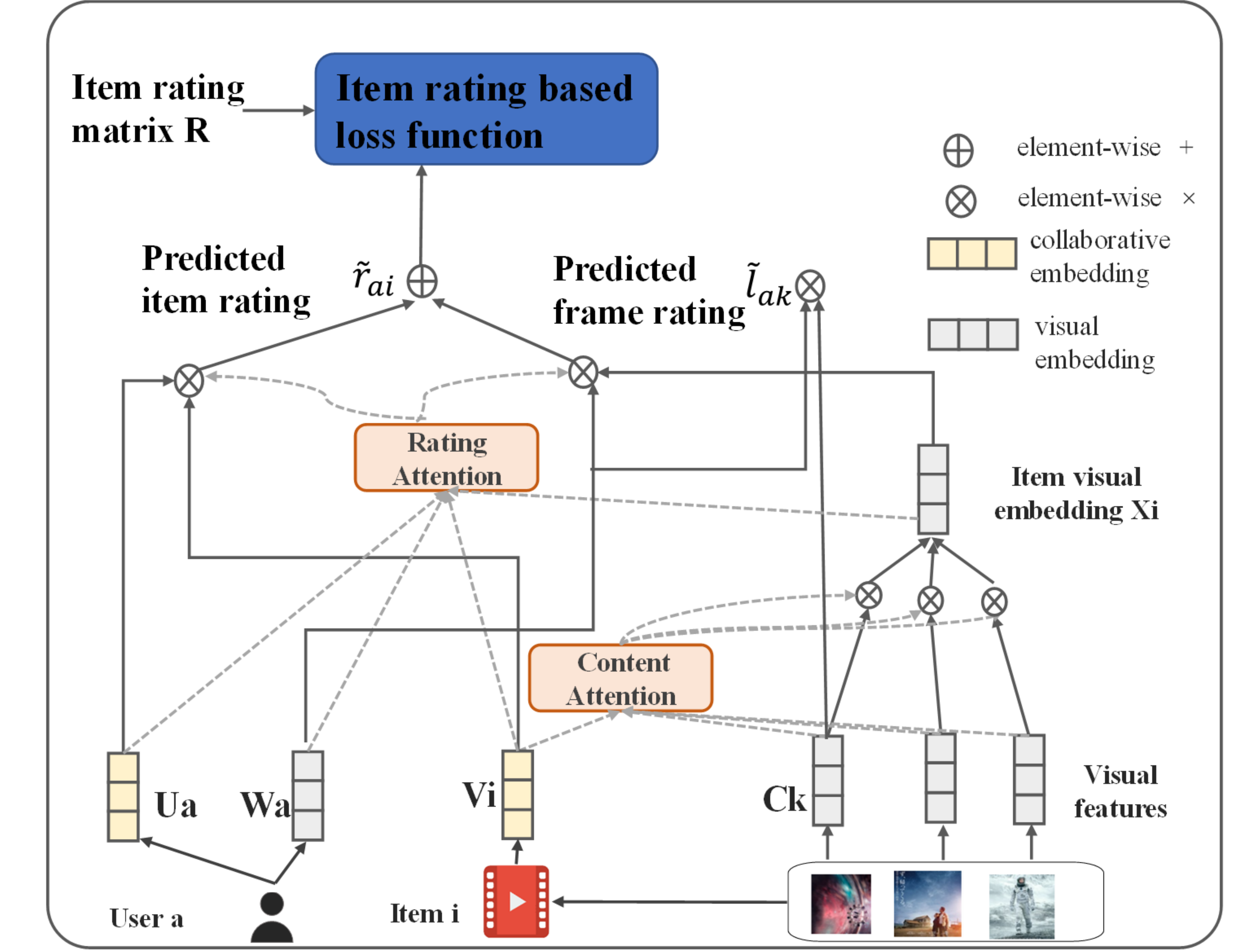}
  \end{center}
  \vspace{-0.4cm}
  \caption{\small{The framework of our proposed JIFR model.} } \label{fig:model1}
  \vspace{-0.2cm}
\end{figure}
\end{small}

\vspace{-0.2cm}
\subsection{The Proposed JIFR Model}
In this part, we illustrate our proposed \emph{J}oint \emph{I}tem and key \emph{F}rame \emph{R}ecommendation~(JIFR) model, with the architecture is show in Fig.~\ref{fig:model1}.
The key idea of JIFR are two carefully designed attention networks for dealing with the content indecisiveness and rating indecisiveness.  Specifically, to tackle the content indecisiveness, the visual content attention attentively learns the visual representation $\mathbf{x}_i$ of each item. By taking the user-item predicted collaborative rating $\mathbf{u}_a^T\mathbf{v}_i$, and the visual content based rating $\mathbf{w}_a^T\mathbf{x}_i$, the rating attention module learns to attentively combine these two kinds of predictions for rating indecisiveness problem.

\emph{Visual Content Attention.} For each multimedia item $i$, the goal of the visual attention is to select the frames from the frameset $S_i$ that are representative for item visual representation. Therefore, instead of simply
averaging all the related images' visual embeddings as the item visual dimension~(Eq.\eqref{eq:vis_avg_i}), we model the item visual embedding as:

\begin{small}
\vspace{-0.2cm}
\begin{equation} \label{eq:item_con}
\mathbf{x}_i=\sum_{j=1}^L \alpha_{ij}s_{ji}\mathbf{P}\mathbf{c}_j,
\end{equation}
\vspace{-0.2cm}
\end{small}

\noindent where $s_{ji}=1$ if image $j$ belongs to the image element of multimedia item $i$.  $\alpha_{ij}$ denotes the attentive weight of image $j$ for multimedia item $i$. The larger the  value of $\alpha_{ij}$,  the more the current visual frame contributes to the item visual content representation.

Since $\alpha_{ij}$ is not explicitly given, we model the attentive weight as a three-layered attention neural network:

\begin{small}
\vspace{-0.2cm}
\begin{equation} \label{eq:att_content}
\alpha_{ij}=\mathbf{w}^1 \times f(\mathbf{W}^1 [\mathbf{v}_i,\mathbf{W}^0\mathbf{c}_j]),
\end{equation}
\vspace{-0.2cm}
\end{small}

\noindent where {\small$\Theta_c=[\mathbf{w}^1,\mathbf{W}^1,\mathbf{W}^0 ]$} is the parameter set in this three layered attention network, and $f(x)$ is a non-linear activation function. Specifically, {\small$\mathbf{W}^0$} is a dimension reduction matrix that transforms  the original visual embeddings~(i.e., {\small$\mathbf{c}_j\in \mathbb{R}^{4096}$}) in a low dimensional visual space.

Then, the final attentive upload history score $\alpha_{ji}$ is obtained by normalizing the above attention scores as:

\begin{small}
\vspace{-0.2cm}
\begin{equation} \label{eq:att_content_nor}
\alpha_{ij}=\frac{exp(\alpha_{ij})}{\sum_{k=1}^L s_{ki}exp(\alpha_{ak})}.
\end{equation}
\vspace{-0.2cm}
\end{small}

\emph{Hybrid Rating Attention.} The hybrid rating attention part models the importance of the collaborative preference and the content based preference for users' final decision making as:

\begin{small}
\vspace{-0.2cm}
\begin{equation} \label{eq:pred_hybrid1}
\hat{r}_{ai}=\beta_{ai1}\mathbf{u}^T_a\mathbf{v}_i+\beta_{ai2}\mathbf{w}^T_a\mathbf{x}_i,
\end{equation}
\vspace{-0.2cm}
\end{small}

\noindent where the first part models the collaborative predicted rating, and the second part denotes the user's visual preference for items. $\beta_{ai1}$ and $\beta_{ai2}$ are the weights that balance these two effects for the user's final preference.

As the underlying reasons for users to balance these two aspects are unobservable, we model the hybrid rating attention as:

\begin{small}
\vspace{-0.2cm}
\begin{flalign} \label{eq:att_huv}
\beta_{ai1}=\mathbf{w}^2 \times f(\mathbf{W}^2[\mathbf{u}_a, \mathbf{v}_i]), \quad
\beta_{ai2}=\mathbf{w}^2 \times f(\mathbf{W}^2[\mathbf{w}_a, \mathbf{x}_i]).
\end{flalign}
\vspace{-0.2cm}
\end{small}

Then, the final rating attention values $\beta_{ai1}$  and $\beta_{ai2}$ are also normalized as:

\begin{small}
\vspace{-0.2cm}
\begin{flalign} \label{eq:att_hybrid_nor}
\beta_{ai1}=\frac{exp(\beta_{ai1})}{ exp(\beta_{ai1})+exp(\beta_{ai2})}=1-\beta_{ai2}.
\end{flalign}
\vspace{-0.2cm}
\end{small}

\textbf{Model Learning and Prediction.} With the two proposed attention networks, the optimization function is the same as Eq.~\eqref{eq:loss_vbpr}. To optimize the objective function, we implement the model in TensorFlow~\cite{abadi2016tensorflow} to train model parameters with mini-batch Adam, which is a stochastic gradient descent based optimization model with adaptive learning rates. For the user-item interaction behavior, we could only observe the positive feedbacks with huge missing ratings. In practice, similar as many implicit feedback based optimization approaches, in each iteration, for each positive feedback, we randomly sample 10 missing feedbacks as pseudo negative feedbacks in the training process~\cite{sigir2017attentive,TKDE2017modeling}. As in each iteration, the pseudo negative feedbacks change with each missing record gives very weak negative signal.

After the model learning process, the multimedia recommendation could be directly computed as in Eq.\eqref{eq:pred_hybrid1}. For each recommended item, as users and images are also learned in the visual dimensions, the key frame recommendation could be predicted as in Eq.\eqref{eq:pred_img}.

\begin{figure*} [htb]
  \begin{center}
      \subfigure[\vspace{-1.5cm}HR@K]{\includegraphics[width=80mm,height=40mm]{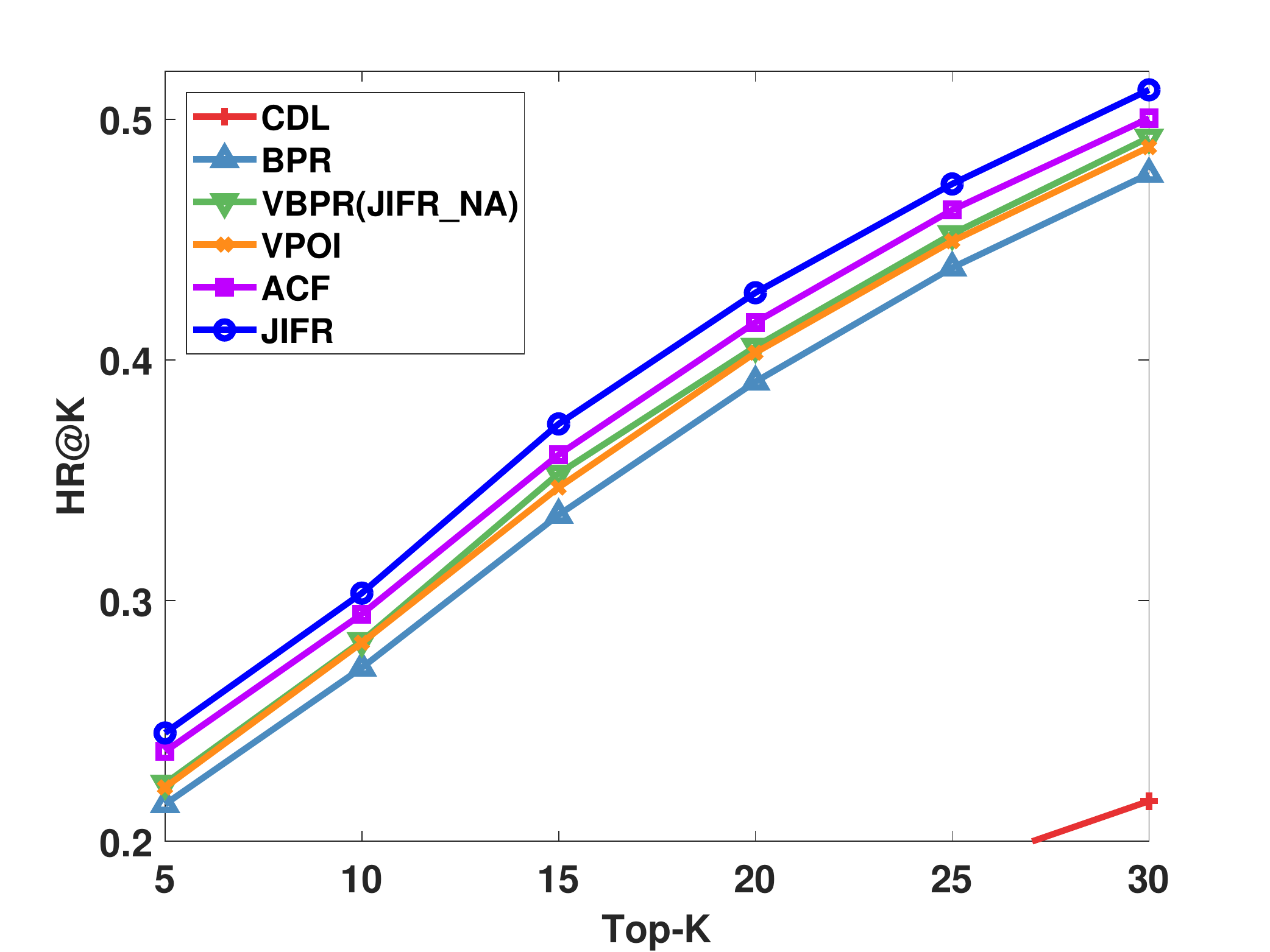} \label{fig:hit_item}}
      \subfigure[\vspace{-1.5cm}NDCG@K]{\includegraphics[width=80mm,height=40mm]{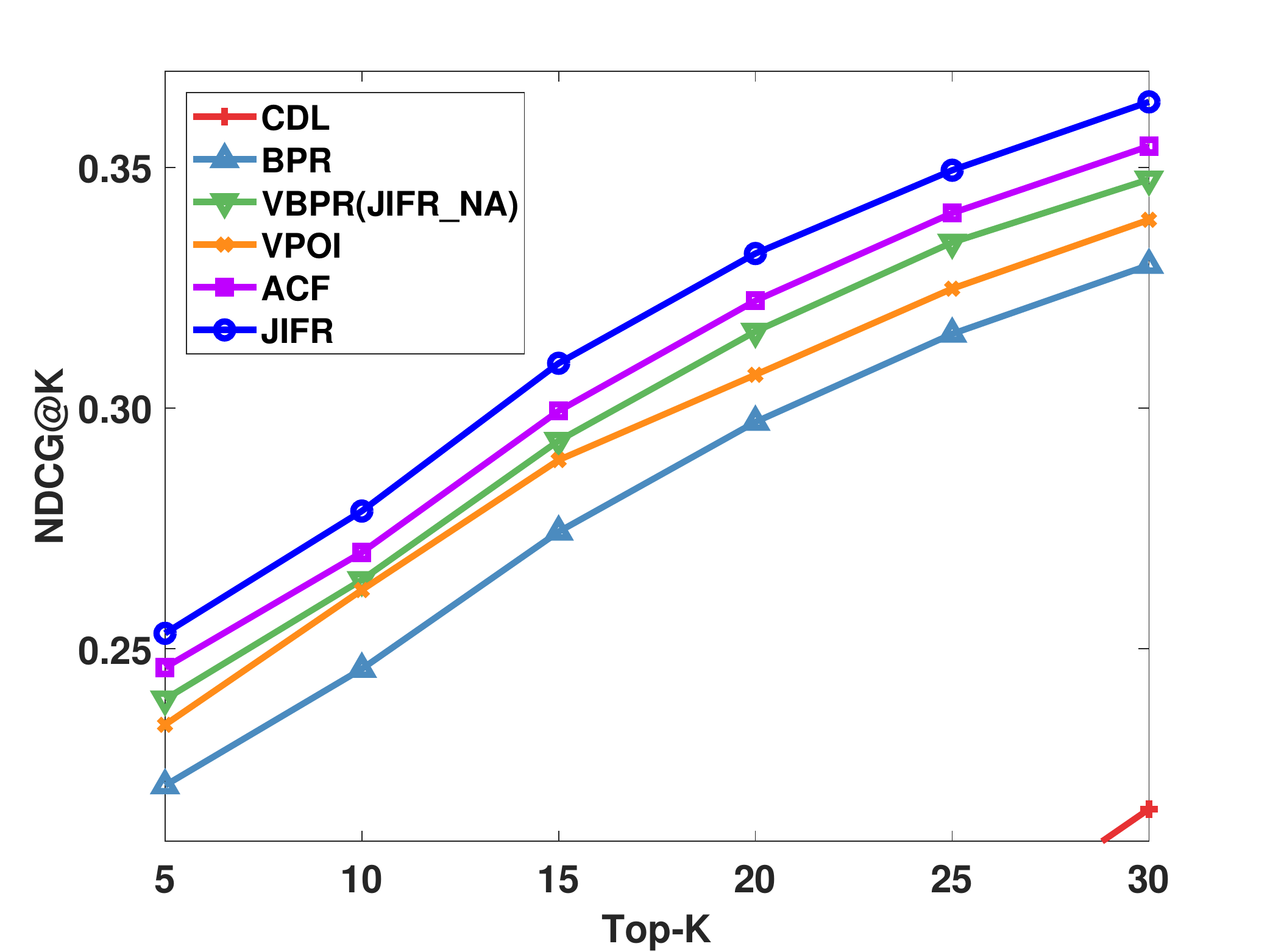}  \label{fig:ndcg_item}}
  \vspace{-0.5cm}
  \caption{Item recommendation performance (Better viewed in color). } \label{fig:ov_item}
  \vspace{-0.4cm}
\end{center}
\end{figure*}

\begin{figure} [htb]
  \begin{center}
  \vspace{-0.2cm}
      \subfigure[\vspace{-1.5cm}HR@K]{\includegraphics[width=40mm,height=25mm]{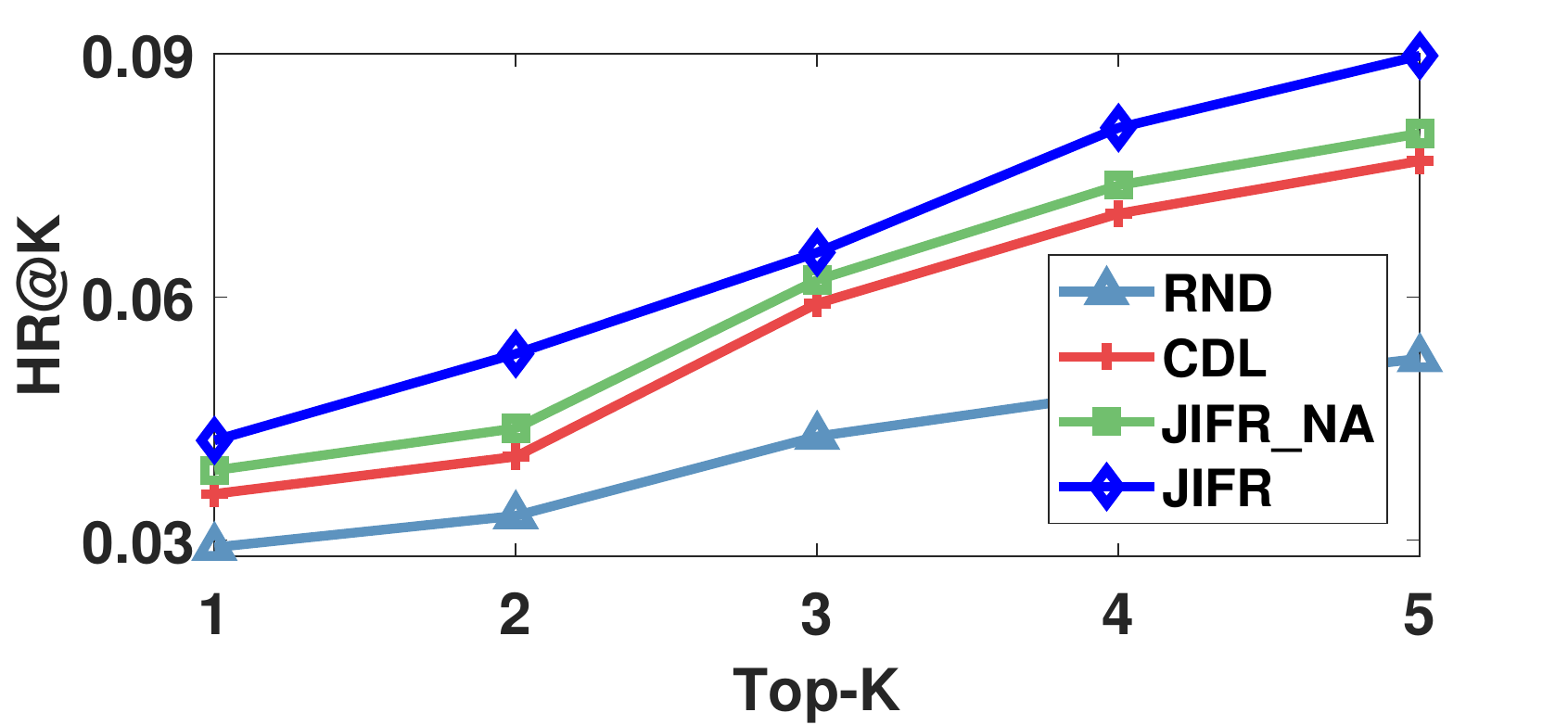} \label{fig:hit_frame}}
      \subfigure[\vspace{-1.5cm}NDCG@K]{\includegraphics[width=40mm,height=25mm]{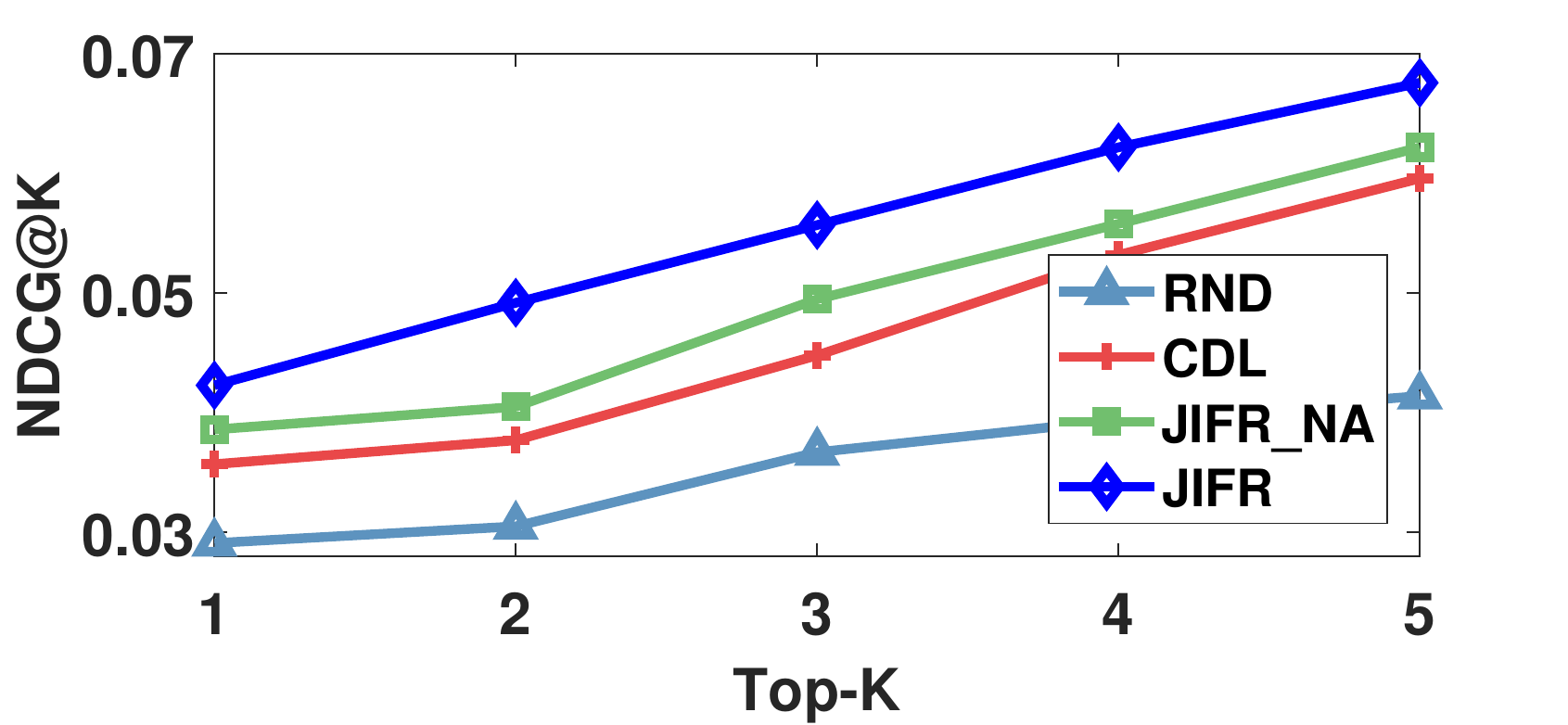}  \label{fig:ndcg_frame}}
  \vspace{-0.5cm}
  \caption{Key frame recommendation performance.} \label{fig:ov_frame}
  \vspace{-0.4cm}
\end{center}
\end{figure}

\subsection{Connections to related models.}
 
\emph{VBPR}~\cite{aaai2016vbpr} extends the classical collaborative filtering model with the additional visual dimensions of users and items~(Eq.\eqref{eq:pred_img}).
By assigning the same weight for all of an items' frames as shown in Eq~\eqref{eq:vis_avg_i}, and without any hybrid rating attention, our proposed item recommendation task degenerates to VBPR.

\emph{ACF}~\cite{sigir2017attentive} is proposed to combine each user's historical rated items and the item components with attentive modeling on top the CF model of SVD++~\cite{koren2009matrix}.  ACF did not explicitly model users in the visual space, and could not be transferred for visual key frame recommendation.

\emph{VPOI}~\cite{WWW2017imagePOI} is proposed for POI recommendation by leveraging the uploaded images of users at a particular POI. In VPOI, each item has a free embedding, and the relationship between images and POIs are used as side information in the regularization terms. 

\emph{KFR}~\cite{sigir2017key} is the one of the first attempts for personalized key frame recommendation.
As KFR relied on users' fine-grained interaction data with frames, it fails when the user-frame interaction data is not available. Besides, KFR is not designed for item recommendation at the same time.

\emph{Attention Mechanism} is also closely related to our modeling techniques. Attention has been widely used in recommendation, such as the importance of historical items in CF models~\cite{TKDE2018nais,sigir2017key}, the helpfulness of review for recommendation~\cite{WWW2018neural,sigir2017key,icjai20183ncf}, the strength of social influence for social recommendation~\cite{SIGIR2018attentive}. In particular, ACCM is an attention based hybrid item recommendation model. It also fuses users and items in the collaborative space and the content space for recommendation~\cite{CIKM2018attention}.  However, the content representation of users and items rely on the features of both users and items. As it is sometimes hard to collect user profiles, this model could not be applied when
users do not have any features, including our proposed problem.

In summary, our proposed model differs greatly from these related models as we perform both joint personalized item and key frame recommendation at the same time. The application scenario has rarely been studied before. From the technical perspective, we carefully design two levels of attentions for dealing with content indecisiveness and rating indecisiveness, which is tailored to discern the visual profiles of users for joint item and key frame recommendation.

\vspace{-0.2cm}
\section{Experiments}
We conduct experiments on a real-world dataset. To the best of our knowledge, there is no public available datasets  with fine-grained user behavior data for evaluating the key frame recommendation performance. To this end, we crawl a large dataset from \emph{Douban}~\footnote{www.douban.com}, which is one of the most popular movie sharing websites in China. We crawl this dataset as for each movie, this platform allows users to show their  preference to each frame of this movie by clicking the ``Like'' button just below each frame.

After data crawling, we pre-process the data to ensure each user and each item have at least 5 rating records.  In data splitting process, we randomly select 70\% user-movie ratings for training, 10\% for validation and 20\% for test.
The pruned dataset has 16,166 users, 12,811 movies, 379,767 training movie ratings, 98,425 test movie ratings, and 4,760 test frame ratings. For each user-item record in the test data, if the user has rated the images of this multimedia item, the correlated user-image records  are used for evaluating the  key frame recommendation performance in the test data. Please note that, to make the proposed model general to the multimedia recommendation scenarios, the fine-grained image ratings in the training data is not used for model learning.

\begin{table}
\begin{scriptsize}
\begin{tabular}{|p{3.8cm}|p{0.75cm}|p{0.55cm}|p{0.55cm}|p{0.6cm}|}\hline
\multirow{3}{*}{Model} & \multicolumn{2}{|c|}{Input} & \multicolumn{2}{|c|}{Task} \\ \cline{2-5}
 & \multirow{2}{*}{Rating} & \multirow{2}{*}{Image} & Item & Frame \\
 & & & Rec& Rec\\ \hline
 BPR~\cite{UAI2009bpr}   & $\surd$  & $\times$  & $\surd$ & $\times$ \\ \hline
 CDL~\cite{CVPR2016comparative} &$\surd$   & $\surd$   & $\surd$ & $\surd$  \\ \hline
 VBPR~\cite{aaai2016vbpr} & $\surd$ &$\surd$  & $\surd$ &$\times$  \\ \hline
 VPOI~\cite{WWW2017imagePOI} & $\surd$ &$\surd$  & $\surd$ &$\times$  \\ \hline
 ACF~\cite{sigir2017attentive} & $\surd$ &$\surd$  & $\surd$ &$\times$  \\ \hline
 \emph{JIFR\_NA}&$\surd$ &$\surd$  &$\surd$ &$\surd$  \\ \hline
 \emph{JIFR}&$\surd$ &$\surd$  &$\surd$ &$\surd$  \\ \hline
\end{tabular}
\vspace{-0.2cm}
\caption{The characteristics of the models.}\label{tab:model}
\end{scriptsize}
\vspace{-0.4cm}
\end{table}

\begin{figure*} [htb]
  \begin{center}
  \vspace{-0.2cm}
\includegraphics[width=150mm]{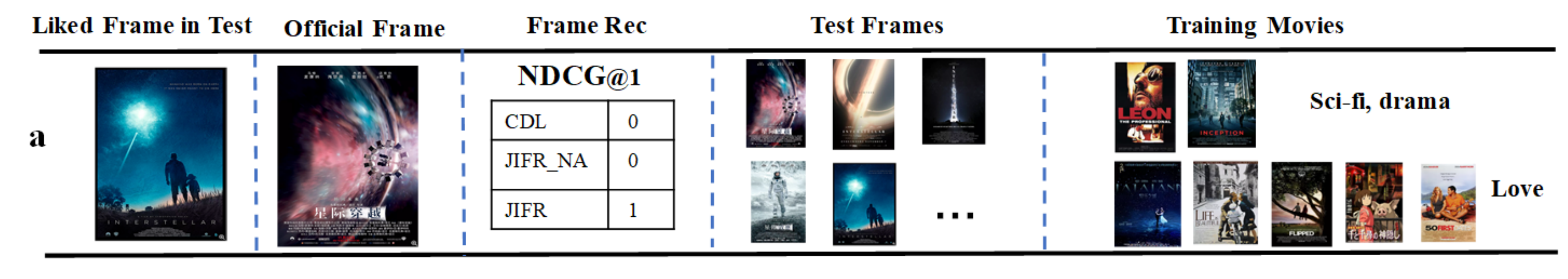}
  \vspace{-0.5cm}
  \caption{\small{Case study of the key frame recommendation of \emph{Interstellar} for a typical user a. }} \label{fig:case}
  \vspace{-0.6cm}
\end{center}
\end{figure*}

\vspace{-0.2cm}
\subsection{Overall Performance}
We adopt two widely used evaluation metrics to measure the top-K ranking performance: the Hit Ratio~(HR@K) and Normalized Discounted Cumulative Gain~(NDCG@K)~\cite{aaai2016vbpr,SIGIR2018attentive,sigir2017attentive}. In our proposed JIFR model, we choose the collaborative latent dimension size $d1$ and the visual dimension size $d2$ in the set [16,32,64], and find when $d1\!=\!d2\!=\!32$ reaches the best performance. The non-linear activation function $f(x)$ in the attention networks is set as the ReLU function. Besides, the regularization parameter is set in range $[0.0001,0.001,0.01,01]$, and $\lambda_1=0.001$ reaches the best performance.

For better illustration, we summarize the details of the comparison  models in Table~\ref{tab:model}, where the second and third column shows the input data of each model, and the last two columns show whether this model could be used for item recommendation and key frame recommendation. The last two rows are our proposed models, with
JIFR\_NA is a simplified version of our proposed model without any attention modeling.

\textbf{Item Recommendation Performance.}  In the item recommendation evaluation process, as the multimedia item size is large, for each user, we randomly select 1000 unrated items as negative samples, and then mix them with the positive feedbacks to select the top-K ranking items. To avoid bias at each time, the process is repeated for 10 times and we report the average results. Fig.~\ref{fig:ov_item} shows the results with different top-K values. In this figure,
some top-K results for CDL are not shown as the performance is lower than the smallest range values of the y-axis. This is because CDL is a visual content based  model without any collaborative filtering modeling, while the collaborative signals are very important for enhancing recommendation performance. Please note that, for item recommendation, our simplified model JIFR\_NA degenerates to VBPR without any attentive modeling. VPOI and VBPR improves over BPR by leveraging the visual data. ACF further improves the performance by learning the attentive weights in the user's history. Our proposed JIFR model achieves the best performance by explicitly modeling users and items in both the latent space and the visual space with two attention networks. For the two metrics, the improvement of NDCG is larger than HR as NDCG considers the ranking position of the hit items.

\textbf{Key Frame Recommendation Performance.} For key frame recommendation, as the detailed user-image information is not available, the models that relied on the collaborative information fails, including  \emph{BPR}, \emph{VBPR}, \emph{KFR}~\cite{sigir2017key}. We also show a simplified \emph{RND} baseline that randomly selects a frame from the movie frames for evaluation. In the evaluation process, all the related frames of this movie is used as the candidate key frames.  Please note that, the ranking list size $K$ for key frame recommendation is very small as we could only recommend one key frame of each multimedia item, while the item recommendation list could be larger. As observed in Fig~\ref{fig:ov_frame}, all models improve over RND, showing the effectiveness of modeling user visual profile. Among all models, our proposed JIFR model shows the best performance, followed by the simplified JIFR\_NA model. This clearly shows the effectiveness of discerning the visual profiles and the collaborative interests of users with attentive modeling techniques.

\begin{table}[!htbp]
\centering
\begin{small}
\vspace{-0.2cm}
\begin{tabular}{|*{6}{c|}}
\hline
Visual & Rating &\multicolumn{2}{|c|}{Item Rec@15}&\multicolumn{2}{|c|}{Frame Rec@3}\\
\cline{3-6}
Att& Att& HR&NDCG&HR&NDCG\\ \hline
AVG &AVG & / & / &/ &/ \\ \hline
AVG&ATT & 2.25\% &  2.42\%  &4.20\%  &4.35\%   \\  \hline
ATT&AVG & 4.02\%    &4.44\% &8.07\% & 8.62\%   \\  \hline
ATT&ATT & 5.49\% &5.94\%  &9.70\% & 10.53\% \\ \hline
\end{tabular}
\end{small}
\caption{Improvement of attention modeling.}\label{tab:att}
\vspace{-0.2cm}
\end{table}

\vspace{-0.2cm}
\subsection{Detailed Model Analysis}
\textbf{Attention Analysis.}  Table~\ref{tab:att} shows the performance improvement  of different attention networks compared to the average setting, i.e., $\alpha_{ji}=\frac{1}{\sum_{k=1}^C s_{ki}}$ for content attention modeling, and $\beta_{ai1}=\beta_{ai2}$ for rating fusion. The ranking list value $K$ is set as $K=15$ for item recommendation and $K=3$  for key frame recommendation.   As can be observed from this table, both attention techniques improve the performance of the two recommendation tasks.  On average, the visual attention improvement is larger than the rating attention. By combining these two attention networks, the two recommendation tasks achieve the best performance.

\textbf{Frame Recommendation Case Study.} Fig~\ref{fig:case} shows the a case study of the recommended frames for user \emph{a} with the movie \emph{Interstellar}. It is regarded as a sci-fi that describes a team explorers travel through a wormhole in space to ensure humanity's survival. In the meantime, the love between the leading actor, and his daughter also touches the audiences. For ease of understanding, we list the training data of this user in the last column, with each movie is represented with an official poster. In the training data, the liked movies contain both sci-fi and love categories. Our proposed JIFR model could correctly recommend the key frame, which is different from the official poster of this movie. However, the remaining models fail. We guess a possible reason is that, as shown in the four column, most frames of this movie are correlated to sci-fi. As the comparison models could not distinguish the important of these frames, the love related visual frame is overwhelmed by the sci-fi visual frames. Our model tackles this problem by learning the attentive frame weights for item visual representation, and user visual representation from her historical movies.

\vspace{-0.2cm}
\section{Conclusions}
In this paper, we studied the general problem of personalized multimedia item and key frame recommendation in the absence of fine-grained user behavior. We proposed a JIFR model to project both users and items into a latent
collaborative space and a visual space. Two attention networks are designed to tackle the content indecisiveness and rating indecisiveness for better discerning the visual profiles of users. Finally, extensive experimental results  on a real-world dataset clearly showed the effectiveness of our proposed model for the two recommendation tasks.

\section*{Acknowledgments}
This work was supported in part by grants from the National Natural Science Foundation of China( Grant No.  61725203, 61722204, 61602147, 61732008, 61632007), the Anhui Provincial Natural Science Foundation(Grant No. 1708085QF155), and the Fundamental Research Funds for the Central Universities(Grant No. JZ2018HGTB0230).

\bibliographystyle{named}
\bibliography{frame_short}

\end{document}